\documentclass{ap-jnmp}

\usepackage{amsfonts}
\usepackage{amsmath}
\usepackage{amsfonts,amssymb}
\usepackage{color}
\usepackage{slashed}
\usepackage{bm}

\arttype{Letter} 

\copyrightauthor{}

\def\bib{B\kern-.05em{I}\kern-.025em{B}\kern-.08em}
\def\btex{B\kern-.05em{I}\kern-.025em{B}\kern-.08em\TeX}

\title{Classical solutions of a massless Wess-Zumino model}

\author{Marco Frasca}

\address{Via Erasmo Gattamelata, 3\\
00176 Rome (Italy)
\email{marcofrasca@mclink.it}}

\begin{document}

\maketitle
\thispagestyle{empty}

\vphantom{\vbox{%
\begin{history}
\received{(Day Month Year)}
\revised{(Day Month Year)}
\accepted{(Day Month Year)}
\end{history}
}}

\begin{abstract}
We provide a current expansion for the classical equations of motion of the massles Wess-Zumino model. In the low-energy limit, there appears a massive behavior for bosonic degrees of freedom and, at small coupling, the fermion field shows the same mass and supersymmetry is overall preserved. In the limit of the coupling running to infinity, the fermion field displays a different equation of motion, a Nambu-Jona-Lasinio-like one, and we cannot draw an identical conclusion.
\end{abstract}

\keywords{Supersymmetry; Wess-Zumino Model; Classical Solutions.}

\section{Introduction}

Supersymmetry (SUSY) is a mathematical framework showing how a symmetry exists relating bosonic and fermionic degrees of freedom. In order for it to hold in any theoretical formulation, known observed particles should have a partner with the same mass. This is not seen and so one expects that, if it is realized in nature, some mechanism should break this symmetry making superpartners very high in mass.

Some mechanism for supersymmetry breaking have been proposed so far. Spontaneous symmetry breaking has been proposed by Fayet and Iliopolous and by O'Raifeartaigh \cite{Fayet:1974jb,O'Raifeartaigh:1975pr}. In this case, auxiliary fields, $F$ or $D$, have a non-null vacuum expectation value. But these approaches do not apply to the minimal supersymmetric standard model (MSSM) and possible extensions. MSSM was firstly proposed by Georgi and Dimopolous \cite{Dimopoulos:1981zb}. In order to overcome the difficulties with breaking of supersymmetry and obtain low-energy phenomenology, they adopted ``softly'' broken SUSY. The idea is to use explicitly breaking terms in the Lagrangian of the model but granting renormalizability and invariance under electroweak symmetry group. A fundamental result due to Witten shows how an index can be computed for a given supersymmetric theory that immediately shows if symmetry breaking can occur \cite{Witten:1982df}.

Our aim in this letter is to provide a way to solve the classical equations of motion of a supersymmetric model in a regime where the value of the coupling can be varied arbitrarily. We choose the Wess-Zumino model as this is the simplest paradigm of supersymmetry \cite{Weinberg:2000cr}. The approach we use is a series in powers of currents as already devised in \cite{Cahill:1985mh} for quantum chromodynamics but we will keep our computations in a completely classical realm.

The reason why this analysis is relevant is that these classical solutions show an interesting behavior: In a small coupling limit supersymmetry is seen to be preserved why all the fields acquire an identical mass. On the other limit, taking the coupling running to infinity, a Nambu-Jona-Lasinio-like equation is seen to emerge for the fermion field and no conclusion can be drawn about supersymmetry being preserved or not.

\section{Massless Wess-Zumino model}

Massless Wess-Zumino model has the Lagrangian \cite{Weinberg:2000cr}
\begin{eqnarray}
     L &=& \frac{1}{2}(\partial A)^2+\frac{1}{2}(\partial B)^2+\frac{1}{2}\bar\psi i\slashed\partial\psi \nonumber \\
       &&-\frac{1}{2}g^2(A^2+B^2)^2-g(\bar\psi\psi A-\bar\psi\gamma^5\psi B)
\end{eqnarray}
being $\psi$ a Majorana field, $A=A^\dagger$ a scalar field and $B=B^\dagger$ a pseudo-scalar field. This Lagrangian is invariant under supersymmetric variations that can be stated in the form
\begin{eqnarray}
    \delta A(x)&=&\bar\epsilon\psi(x) \nonumber \\
    \delta B(x)&=&\bar\epsilon\gamma^5\psi(x) \nonumber \\
    \delta\psi(x) &=& \partial_\mu(A-\gamma^5B)\gamma^\mu\epsilon.
\end{eqnarray}
These hold on-shell because we have already removed the auxiliary fields using the equations of motion. From Witten index we know that for this model spontaneous symmetry breaking does not occur \cite{Witten:1982df}. Equations of motion are
\begin{eqnarray}
\label{eq:susy}
    \partial^2A+2g^2A(A^2+B^2)&=&-g\bar\psi\psi \nonumber \\
    \partial^2B+2g^2B(A^2+B^2)&=&g\bar\psi\gamma^5\psi \nonumber \\
    i\slashed\partial\psi&=&2g(A-\gamma^5B)\psi.
\end{eqnarray}
Now, let us consider the following two equations
\begin{eqnarray}
\label{eq:susy2}
    \partial^2A+2g^2A(A^2+B^2)&=&j \nonumber \\
    \partial^2B+2g^2B(A^2+B^2)&=&j_5
\end{eqnarray}
being $j=-g\bar\psi\psi$ and $j_5=g\bar\psi\gamma^5\psi$ These imply that $A=A[x;j,j_5]$ and $B=B[x;j,j_5]$ and so we take the functional Taylor series
\begin{eqnarray}
    A &=& A_0(x)+\int dx'\left.\frac{\delta A}{\delta j(x')}\right|_{j,j_5=0}j(x')
    +\int dx'\left.\frac{\delta A}{\delta j_5(x')}\right|_{j,j_5=0}j_5(x')
     \nonumber \\
    &&+\frac{1}{2}\int dx'dx''\left.\frac{\delta^2 A}{\delta j(x')\delta j(x'')}\right|_{j,j_5=0}j(x')j(x'')
    +\int dx'dx''\left.\frac{\delta^2 A}{\delta j_5(x')\delta j(x'')}\right|_{j,j_5=0}j_5(x')j(x'') \nonumber \\
    &&+\frac{1}{2}\int dx'dx''\left.\frac{\delta^2 A}{\delta j_5(x')\delta j_5(x'')}\right|_{j,j_5=0}j_5(x')j_5(x'')+\ldots \nonumber \\
    B &=& B_0(x)+\int dx'\left.\frac{\delta B}{\delta j(x')}\right|_{j,j_5=0}j(x')
    +\int dx'\left.\frac{\delta B}{\delta j_5(x')}\right|_{j,j_5=0}j_5(x')
     \nonumber \\
    &&+\frac{1}{2}\int dx'dx''\left.\frac{\delta^2 B}{\delta j(x')\delta j(x'')}\right|_{j,j_5=0}j(x')j(x'')
    +\int dx'dx''\left.\frac{\delta^2 B}{\delta j_5(x')\delta j(x'')}\right|_{j,j_5=0}j_5(x')j(x'') \nonumber \\
    &&+\frac{1}{2}\int dx'dx''\left.\frac{\delta^2 B}{\delta j_5(x')\delta j_5(x'')}\right|_{j,j_5=0}j_5(x')j_5(x'')+\ldots
\end{eqnarray}
having set $A_0(x)=A[x;0,0]$ and $B_0(x)=B[x;0,0]$. Then we insert this series into eqs.(\ref{eq:susy2}) and we get the leading order equations
\begin{eqnarray}
   \partial^2A_0+2g^2(A_0^2+B_0^2)A_0 &=& 0 \nonumber \\
   \partial^2B_0+2g^2(A_0^2+B_0^2)B_0 &=& 0
\end{eqnarray}
that can be solved exactly \cite{Frasca:2009bc}. This is easy to see as we set $B_0=\eta_pA_0$ with $\eta_p^2=1$ the parity factor. Then, setting $A_0=\phi$, we have to solve the equation $\partial^2\phi+4g^2\phi^3=0$ that has the solution $\phi(x)=\mu(1/2^\frac{1}{4}g^\frac{1}{2}){\rm sn}(p\cdot x+\theta,i)$ being $\mu$ and $\theta$ integration constants and sn a Jacobi elliptic function. This holds provided $p^2=\sqrt{2}\mu^2g$ that can be interpreted as a dispersion relation for a massive particle. At this order both boson fields get a mass but supersymmetry is preserved as we will also see for the Fermion field at small coupling. Next-to-leading order equations give
\begin{eqnarray}
   \partial^2A_1+8g^2\phi^2A_1+4g^2\phi^2\eta_p\bar B_1&=&\delta^4(x) \nonumber \\
   \partial^2\bar A_1+8g^2\phi^2\bar A_1+4g^2\phi^2\eta_pB_1&=& 0 \nonumber \\
   \partial^2B_1+8g^2\phi^2B_1+4g^2\phi^2\eta_p\bar A_1&=& \delta^4(x) \nonumber \\
   \partial^2\bar B_1+8g^2\phi^2\bar B_1+4g^2\phi^2\eta_p A_1&=& 0
\end{eqnarray}
where we have set $A_1=\left.\delta A/\delta j(x)\right|_{j,j_5=0}$, $\bar A_1=\left.\delta A/\delta j_5(x)\right|_{j,j_5=0}$, $B_1=\left.\delta B/\delta j_5(x)\right|_{j,j_5=0}$, $\bar B_1=\left.\delta B/\delta j(x)\right|_{j,j_5=0}$. Now, introducing the Green function $\partial^2\Delta+8g^2\phi^2\Delta=\delta^4(x)$, this set of equations can be turned into a single integral equation by noting that $B_1=A_1=\phi_1$ and so, we get
\begin{equation}
\label{eq:phi1}
   \phi_1(x)=\Delta(x)+16g^4\int d^4x'\Delta(x-x')\phi^2(x')\int d^4x''\Delta(x'-x'')\phi^2(x'')\phi_1(x'').
\end{equation}
Use has been made of the equations
\begin{eqnarray}
   \bar A_1 &=& -4g^2\int d^4x'\Delta(x-x')\phi^2(x')\eta_pB_1 \nonumber \\
   \bar B_1 &=& -4g^2\int d^4x'\Delta(x-x')\phi^2(x')\eta_pA_1
\end{eqnarray}
that can be given by $\phi_1$. Now we turn to the computation of the Green function $\Delta(x)$. This can be obtained immediately \cite{Frasca:2009nt} as
\begin{equation}
\label{eq:prop}
    \Delta(p^2)=\sum_{n=0}^\infty\frac{B_n}{p^2-m_n^2+i\epsilon}
\end{equation}
being
\begin{equation}
    B_n=(2n+1)^2\frac{\pi^3}{4K^3(i)}\frac{e^{-(n+\frac{1}{2})\pi}}{1+e^{-(2n+1)\pi}}.
\end{equation}
and $m_n=(2n+1)(\pi/2K(i))(4/3)^\frac{1}{4}g^\frac{1}{2}\mu$ a possible mass spectrum when this propagator enters in quantum field theory and $K(i)$ an elliptic integral $\int_0^\frac{\pi}{2}1/\sqrt{1+\sin^2\theta}d\theta$ and the phase of $\phi$ is chosen to be $K(i)$. Eq.(\ref{eq:phi1}) admits a simple solution in the limit of very low momenta. In this case we get the simple expression
\begin{equation}
   \Delta_0(x-x')=-\frac{\sqrt{3}}{4\mu^2g}\delta^4(x-x')
\end{equation}
proper to a contact interaction. So, putting this into eq.(\ref{eq:prop}) one gets
\begin{equation}
   \phi_1(x)=\frac{\sqrt{3}}{2\mu^2g}\delta^4(x)
\end{equation}
and so,
\begin{equation}
   \bar A_1 = \bar B_1 = \eta_p\frac{3}{2\sqrt{2}}\frac{1}{\mu^2g}\delta^4(x)
\end{equation}
Finally, the low-energy series for our solution is given by
\begin{eqnarray}
    A &=& A_0(x)+\frac{\sqrt{3}}{2}\frac{1}{\mu^2g}\left(j(x)
    +\sqrt{\frac{3}{2}}\eta_pj_5(x)\right)+O(j^2)+O(j^2_5)+O(jj_5) \nonumber \\
    B &=& B_0(x)+\frac{\sqrt{3}}{2}\frac{1}{\mu^2g}\left(j_5(x)+\sqrt{\frac{3}{2}}\eta_pj(x)\right)+O(j^2)+O(j^2_5)+O(jj_5) 
\end{eqnarray}
Now, turning our attention to the fermion field, we get
\begin{equation}
   i\slashed\partial\psi-2g\phi(1-\eta_p\gamma^5)\psi=
   -\sqrt{3}\frac{g}{\mu^2}\left(1-\sqrt{\frac{3}{2}}\eta_p\gamma^5\right)(\bar\psi\psi+\gamma^5\bar\psi\gamma^5\psi)\psi.
\end{equation}
Looking at the limit $g\rightarrow 0$, we observe that Fermion field has the same mass of the bosonic degrees of freedom. In this case one gets the equation of motion
\begin{equation}
  i\slashed\partial\psi_c-2g\phi(1-\eta_p\gamma^5)\psi_c=0.
\end{equation}
This equation can be solved by putting $\psi_c=e^{-ikS(x)}u_{ps}$, being $k$ a real constant, and we get $k\gamma\cdot\partial Su_{ps}-2g\phi(1-\eta_p\gamma^5)u_{ps}=0$. But we notice that $\phi$ depends just on the product $\xi=p\cdot x$ and we choose $\partial_\xi S(\xi)=2g\phi(\xi)$. So, we are left with the algebraic equation $[\slashed p-k^{-1}(1-\eta_p\gamma^5)]u_{ps}=0$. We solve this equation by remembering that $p^2=\sqrt{2}\mu^2g$ and we get our proof: In the small coupling limit supersymmetry is preserved at classical level as also happens in quantum field theory in agreement with Witten index for this theory.

This equation appears interesting instead when the limit $g\rightarrow\infty$ is taken. Then, at the leading order, the contribution coming from the external field $\phi$ drops out with respect to the non-linear one as one has $\phi\sim 1/\sqrt{g}$ and an overall $\sqrt{g}$ with respect the latter going like $g$. So, one gets
\begin{equation}
   i\slashed\partial\psi_0=
   -\sqrt{3}\frac{g}{\mu^2}\left(1-\sqrt{\frac{3}{2}}\eta_p\gamma^5\right)(\bar\psi_0\psi_0+\gamma^5\bar\psi_0\gamma^5\psi_0)\psi_0
\end{equation}
that is a Nambu-Jona-Lasinio-like equation of motion. Working in quantum field theory, this could give rise to a symmetry breaking but already at classical level we observe a fermion with a different mass from the boson fields. Supersymmetry appears to be broken in classical equations of motion but we cannot exclude that quantum corrections can change the situation in agreement with Witten index.

\section{Conclusions}

We have provided an approach to solve the classical equations of motion for the Wess-Zumino model. This is a series in powers of currents. The solutions we obtained can be analyzed both at small and large coupling showing different behavior for the fermion field. In the former case, supersymmetry appears to be preserved as both bosonic and fermionic degrees of freedom acquire the same mass. This does not happen in the limit of the coupling running to infinity. In this case a Nambu-Jona-Lasinio-like equation is obtained and no conclusion about the mass of the fermion field can be drawn unless we turn to a quantum field theory.

\end{document}